\documentclass[9pt]{article}
\usepackage{latexsym}
\usepackage{graphicx}
\usepackage{amsfonts,amssymb}

\begin{document}
\title{\textbf{A New Comparative Definition of Community and Corresponding Identifying Algorithm}}
\author{Yanqing Hu, Hongbin Chen, Peng Zhang, Menghui Li, Zengru Di, Ying Fan\footnote{Author for correspondence: yfan@bnu.edu.cn}\\
\\\emph{Department of Systems Science, School of Management,}\\
\emph{Center for Complexity Research,}\\
\emph{Beijing Normal University, Beijing 100875, P.R.China}}
\maketitle
\begin{abstract}
In this paper, a new comparative definition for community in
networks is proposed and the corresponding detecting algorithm is
given. A community is defined as a set of nodes, which satisfy that
each node's degree inside the community should not be smaller than
the node's degree toward any other community. In the algorithm, the
attractive force of a community to a node is defined as the
connections between them. Then employing attractive force based
self-organizing process, without any extra parameter, the best
communities can be detected. Several artificial and real-world
networks, including Zachary Karate club network and College football
network are analyzed. The algorithm works well in detecting
communities and it also gives a nice description for network
division and group formation.
\end{abstract}



\section {Introduction}
Many physicists have become interested in the study of networks
describing the topologies of wide variety of systems\cite{1,Newman
reviwe,new reviwe}, such as the world wide web\cite{www}, social and
communication networks\cite{social network1,social network2},
biochemical networks\cite{metabolism} and many more. Many networks
are found to divide naturally into communities. Nodes belonging to a
tight-knit community are more than likely to have other properties
in common. In the world wide web, community analysis has uncovered
thematic clusters. In biochemical or neural networks, communities
may be functional groups. As a result, the problem of identification
of communities has been the focus of many recent efforts. Many
different algorithms are proposed\cite{3,4,6,9,10,12,14,18,GN,Newman
fast algorithm,Newman mixture algorithm,2005 pans community
defination,Lshell,EO,Potts}(see \cite{4} as a review).

Communities within networks can loosely be defined as subsets of
nodes which are more densely linked, when compared to the rest of
the network. Modularity $Q$ \cite{Newman evaluating algorithm} was
presented as a index of community structure and now has been widely
accepted \cite{4,14,17,EO} as a measure for the communities.
Modularity $Q$ was introduced by Newman and Girvan as follow:
\begin{equation}Q=\sum_{r}{(e_{rr}-a_{r}^{2})}\label{Q}\end{equation}
where $e_{rr}$ is the fraction of links that connect two nodes
inside the community $r$, $a_{r}$ is the fraction of links that have
one or both vertices in side the community $r$, and sum extends to
all communities $r$ in a given network. Note that this index
provides a quantitative measurement to decide the best division of
network. The larger the value of $Q$, the more accurate is a
partition into communities. So maximizing modularity $Q$ can also
detect communities. Actually, there are already many algorithms of
maximizing $Q$ such as Extremal Optimization (EO) \cite{EO}, Greedy
algorithm \cite{10} and other optimal algorithms. There are also
many other algorithms to identify communities in complex networks
such as GN algorithm \cite{GN,Newman evaluating algorithm}, random
walks method \cite{6}, edge clustering coefficient method \cite{3},
and spectral analysis\cite{3}. When the methods can only produce the
dendrogram of the community structure, the best partition is usually
obtained by maximizing modularity $Q$. Unfortunately, modularity $Q$
maximization problem was proved to be a NPC problem \cite{Q NPC}.
Moreover, it has been proved that modularity $Q$ measurement may
fail to identify modules smaller than a scale which depends on the
total number $L$ of links of the network and on the degree of
interconnectedness of the modules, even in cases where modules are
unambiguously defined \cite{23}.

There are also other community definitions based on the topology of
networks, such as self-referring definitions and comparative
definitions. The basic self-referring definition is a clique,
defined as a subgroup of a graph containing more than two nodes
where all the nodes are connected to each other by means of links in
both directions. In other words, this is a fully connected subgraph.
This is a particularly strong definition and rarely fulfilled in
real sparse networks for larger groups \cite{clique}. The another
referring community definition is k-core which is defined as a
subgraph in which each node is adjacent to at least a minimum
number, $k$, of the other nodes in the subgraph. It is weaker than
clique but it is very hard to find the optimal $k$ when we want to
detect the best partition of the network. Comparative definitions
are given on the basis of links comparison. There are three kinds of
comparative definitions which are called LS-set, strong and weak
community definition. LS-set is defined as a set of nodes in which
each of its subsets has more ties to its components within the set
than outside \cite{social network}. The LS-set definition is also
quite stringent. Moreover, it is a very tough problem to detect all
the LS-sets in a network. In order to relax the constraints,
Raddichi et al. \cite{2005 pans community defination} proposed the
strong definition and weak definition. In a strong community, each
node has more connections within the community than with the rest of
the network and in a weak community the sum of all degrees within
the community is larger than the sum of all degrees toward the rest
of the network. Based on these comparative definitions, the
self-contained algorithm is developed, which is similar with GN
algorithm for finding strong or weak communities in a network. But
it is very costly.

In this article, following the basic idea of comparative
definitions, we define community as: a community is a set of nodes,
each node's degree inside the community should be bigger than or at
least equal to its degree link to any other community. This
definition is different from other comparative definitions. The
strong, weak and LS-sets definitions are presented by comparing the
degree in the community with the degree in the whole rest network.
But our community definition is designed by comparing the degree in
the community with the degree in each rest community, not the whole
rest network.

Then how to detect the communities in a network based on our
definition? Obviously, whether a node belongs to a community or not
is determined by its connections. We can define the attractive force
of a community to a node by the links connect them. Employing
attractive force based self-organize process, we can detect
community structures without any extra parameter. The algorithm also
gives a nice description of the affection of a community to a node
and group formation process. With the formation of communities,
individual will choose and change its position according to its
friends continuously until the partition become clear.

This paper is organized as follows. Section 2 gives our comparative
definition for communities in networks. Then in Section 3, the
corresponding algorithm is given in details. The application of the
definition and the algorithm in \emph{ad hoc} networks, Zachary
karate club network, and College football network are presented in
Section 4. Some concluding remarks are put in Section 5.

\section{Quantitative Definitions of Community}
\subsection{Previous comparative definition}
The most important comparative definitions of community are strong
and weak definitions, which are proposed by Raddichi \emph{et
al}.\cite{2005 pans community defination}. Suppose there is a
network $G$ which has $n$ nodes and it can be represented
mathematically by an adjacency matrix $A$ with elements $A_{i,j}=1$
if there is an edge from $i$ to $j$ and $A_{i,j}=0$ otherwise.

\textbf{Definition of Community in a Strong Sense}. The subnetwork
$V$ is a community in a strong sense if for any $i$ belonged to $V$
we have
\begin{equation}\sum_{j\in V}A_{i,j}>\sum_{j\in(G-V)}A_{i,j}\label{strong definition}
\end{equation}

\textbf{Definition of Community in a Weak Sense}. The subnetwork $V$
is a community in a weak sense if we have
\begin{equation}\sum_{i,j\in V}A_{i,j}>\sum_{i,\in V,j\in(G-V)}A_{i,j}\label{weak definition}
\end{equation}
Obviously, strong community definition concerns the situation of
every node, but the weak sense takes a community as a whole. From
the strong (weak) definition of community we can easily get that if
$V_{1},V_{2}\subseteq G$ satisfy strong (weak) definition then we
have $V_{1}\bigcup V_{2}$ also satisfy strong (weak) definition.
Raddichi \emph{et al}.\cite{2005 pans community defination} call
this phenomena as self-contained and use self-contained algorithm to
detect communities, which is similar with GN algorithm for finding
strong or weak communities in a network. But it is very costly.

\subsection{New community definition}
Inspired by the above strong and weak definitions, we define the
community as follow.

\textbf{Definition of Community:} if $V_{1},V_{2},\cdots,V_{m}$ are
$m$ communities of $G$, $V_{k}, k=1,2,\cdots,m$ should satisfy that
\begin{equation}\bigcup_{k=1}^{k=m}V_{k}=G\end{equation}
 and
\begin{equation}
\forall\,\,j\in\,\,V_{k},\ \sum_{i\in\,V_{k}}A_{i,j}\geq
max\{\sum_{i\in\,V_{t}}A_{i,j},t=1,2,\cdots,m\}\label{new
definition}
\end{equation}
This definition can be summarized as: a community should satisfy
that each node's degree inside the community should not be smaller
than the node's degree toward any other community. The same as the
strong sense, our definition also focus on the situation of node.
But different from comparing the degree in the community with the
degree in the whole rest network, our definition compare the degree
in the community with the degree in each rest community instead of
the whole rest network. Obviously, our definition is weaker than the
strong definition. Here we can also give an another most weak
community definition: in a community, the sum of all degree inside
the community should not be smaller than the sum of degree toward
any one other community. The same as the weak sense, our most weak
community definition focus on the case of community instead of the
single node. The difference between the weak definition and our the
most weak definition is that the weak definition compare the sum of
degree inside the community with the sum of degree towards the whole
rest network, but the most weak one compares the sum of degree
inside the community with the sum of degree towards any other
community. In the following discussion, we only deal with the new
definition given by formula (\ref{new definition}).

\section{Algorithm}
In order to detect the community structure under our new definition,
we set each node and its random half of neighbors to be a community
initially. Then we define the attractive force by the connections
among nodes and let the communities be self-organized with the
forces. When the community structure become fixed, the survivors
will be the best partitions which satisfy the above definition
naturally.

Let $F_{k,i}$ denotes the attractive force of community $k$ to node
$i$ and $F_{k,i}$ can be calculate out by the formula
\begin{equation}F_{k,i}=\sum_{j\in V_{k}}A_{i,j}\end{equation}
Then our algorithm is defined as follows.
\begin{enumerate}
\item We initially set each node and its random half neighbors to be a community.
If a node has $h$ neighbors and $h$ is odd, we let the node and it's
random $\frac{h+1}{2}$ nodes as a community. If two or more than two
communities are the same, just keep one of them. So after the first
step the network is partitioned to $n$ or less than $n$ overlapping
communities. $n$ is the number of nodes in the network.

\item Calculate $F_{k,i}$ for all $k$ and $i$.

\item For every node, move it into the community or communities
with the largest attractive force respectively at the same time.

\item Check all communities, if two or more than two communities
are the same, just keep one of them.

\item Repeat step $2$ to step $4$ until sufficient $N$ steps or the partition be fixed.
\end{enumerate}

The time complexity of our algorithm is $O(n^{2})$. Step $1$ runs
in time $O(dn)$, step $2$ in $O(n^{2})$, step $3$ in $O(n^{2})$,
step $4$ in $O(n^{2})$ and the repeated time in step $5$ is
uncertain, where $d$ is the average degree. According to the
numerical experiments in artificial networks, around $10$
repeating steps, the partition will be fixed. So we think the time
complexity is $O(n^{2})$. It is lower than many algorithms for
detecting community structures.

Even our definition of communities is not a self-contained one as
strong and weak definitions, there should be more than one
partitions that may satisfy our community definition. So we keep
some stochastic factors in our initial partition and run the
algorithm several times. Then we could report the average result
or choose the best one from all the partitions. Here we introduce
another indicator for evaluating the partitions. We think the best
partition should satisfy that there are more connections inside
the communities and less connections outside the communities. So
we use the proportion of average connection density inside the
communities and the connection density outside the communities to
measure how reasonable a partition is. This kind of measurement
can be defined as following. Suppose the network contains $n$
nodes and $L$ connections and is partitioned to $m$ communities.
$n_{i}, i=1,2\cdots,m$ denotes the number of nodes in the $i$th
community and $L_{i}, i=1,2\cdots,m$ denotes the number of
connections in the $i$th community. Then the average connection
density inside the communities is
\begin{equation}
D_{in}=\frac{2\sum_{i=1}^{m}{L_{i}}}{\sum_{i=1}^{m}{n_{i}^{2}}-n},
\end{equation}
and the connection density outside the communities
\begin{equation}
D_{out}=\frac{2L-2\sum_{i=1}^{m}{L_{i}}}{n^{2}-\sum_{i=1}^{m}{n_{i}^{2}}}.
\end{equation}
Then the measurement $H$ can be defined as
$H=\frac{D_{in}}{D_{out}}$ and when there only one community $H=0$.
Obviously, larger $H$ means more reasonable partition.

\section{Application in \emph{ad hoc} and Real Networks}
\subsection{Algorithm on artificial networks}
\begin{figure}
\center
\includegraphics[width=8cm]{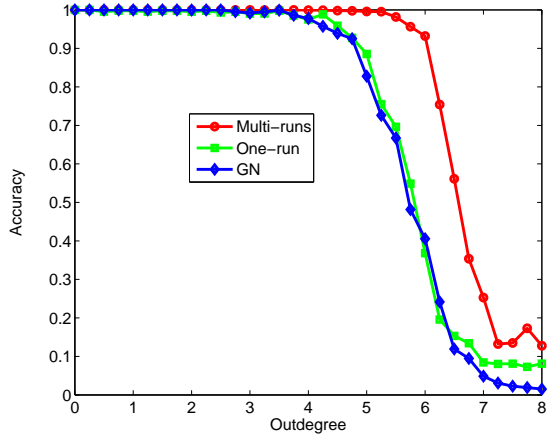}
\caption{The accuracies of our algorithms and GN algorithm. From the
plot we can see that the accuracy of one-run algorithm is similar
with GN algorithm. The best partition of the multi-runs with the aid
of indicator $H$ is better than GN algorithm when the out degree
becomes larger. Where we run $15$ times for each network for
multi-runs. Each point is the average of $20$ realizations of
networks. }\label{artificial_network}
\end{figure}

In order to test our algorithm, we apply it on computer-generated
random networks with a well-known predetermined community structures
and some real networks. The accuracy of the algorithm is evaluated
by similarity function $S$ \cite{Ying_Fan}. Each network has $n=128$
nodes divided into $4$ communities of $32$ nodes each. Edges between
two nodes are introduced with different probabilities depending on
whether the two nodes belong to the same community or not: every
node has $\langle k_{intra}\rangle$ links on average to its fellows
in the same community, and $\langle k_{inter}\rangle$ links to the
outer-world, keeping $\langle k_{intra}\rangle+\langle
k_{inter}\rangle=16$. For each given out degree $\langle
k_{inter}\rangle$, we produce 20 realizations of networks. Then for
each network, we first run the algorithm one time and give the
average accuracy of 20 networks as One-run shown in
Fig.\ref{artificial_network}. Then we run the algorithm 15 times for
each network and choose the best partition with the aid of indicator
$H$. The average accuracy of 20 networks is also shown as Multi-runs
in Fig.\ref{artificial_network}. Comparing our algorithm with GN
algorithm \cite{GN,Newman evaluating algorithm}, we could find that
the accuracy of One-run algorithm is similar with GN and the
accuracy of Multi-runs algorithm is better than GN. Moreover, GN
algorithm need an extra index $Q$ and the time complexity is high,
but our algorithm do not need any extra parameters and has lower
time complexity.

\begin{figure}
\center
\includegraphics[width=8cm]{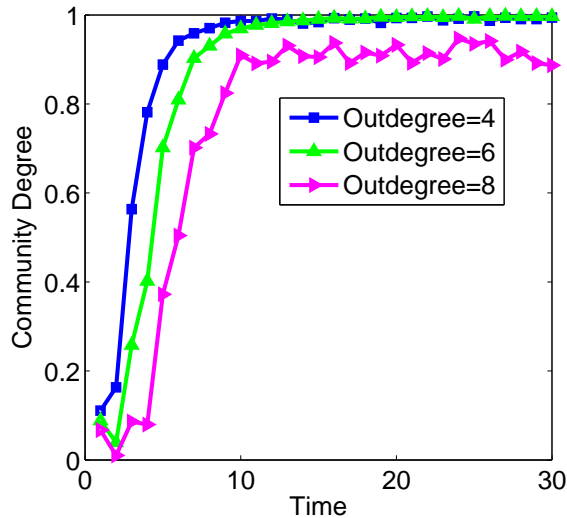}
\caption{The evolution of community degree with the process of the
algorithm. The results are for one-run algorithm. We can see that
when the community structure is no very fuzzy, one-run algorithm
services our community definition very well.}\label{Commdegree}
\end{figure}

We also test that with the process of our algorithm, to what extent
the partition satisfies our definition. For a given partition
$V_{i}$, $i=1,2,\cdots,m$, we define its community degree $CD$ as
the ratio:
\begin{equation}
CD=\frac{\sum_{i=1}^{m}{|\tilde{V}_{i}|}}{\sum_{i=1}^{m}{|V_{i}|}},
\end{equation}
where $\tilde{V}_{i}$ denotes the subset of $V_{i}$, in which each
node satisfy the requirement of our definition for community, that
is node's inter degree is larger or equal to its intra degree
between any other community. The numerical experiments results tell
us when the community structure is not very fuzzy, the algorithm
will finally produce a partition that satisfy our definition very
well. The community degree tends to 1. When the community structure
is very fuzzy, it is hard to find the partition that satisfy the
definition exactly.

\begin{figure}
\center
\includegraphics[width=6cm]{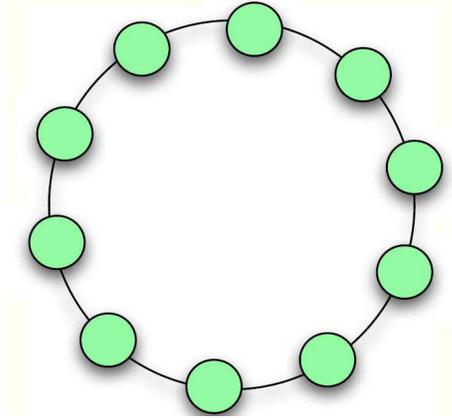}
\caption{The circles represent the communities in which each pair of
nodes are connected. The circles be connected to each other by the
minimal number of links. The plot is cited from
\cite{23}}\label{circle}
\end{figure}

Furthermore, recently Santo Fortunato and Marc Barthelemy \cite{23}
proved that modularity $Q$ may fail to identify small communities
and give a kind of network as shown in Fig.\ref{circle}. We test our
algorithm on this kind of networks. When each circle contains a
clique with $3$ or more than $3$ nodes, our algorithm can detect all
the pre-determinate communities (circles) always.

\subsection{Zachary karate club network}
\begin{figure}
\center
\includegraphics[width=6cm]{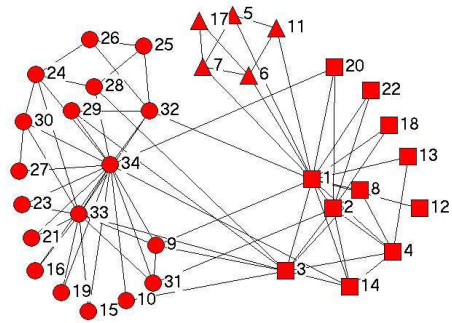}
\caption{The community structure of Zachary Karate club network. Our
algorithm detects $3$ communities which are depicted by circles,
squares and triangles. When we set $1,2,33,34$ as leaders, the
partition is the same. But if we set $1,33,34$ as leaders, the
network will be divided into $2$ communities. Circles represent a
community and the rest is another one, which corresponds to the
actual division.}\label{ZK}
\end{figure}

When apply our algorithm to real network, first we use the popular
Zachary karate club network\cite{Origion ZK}, which is considered as
a simple workbench for community finding
methodologies\cite{GN,Newman fast algorithm,Newman mixture
algorithm,Systematic,Newman evaluating algorithm, Lshell}. This
network was constructed with the data collected observing $34$
members of a karate club over a period of $2$ years and considering
friendship between members. By our algorithm, $3$ communities are
detected (as shown in Fig. \ref{ZK}). The partition is reasonable
compared with the actual division of the club members.

As mentioned above, there may be many partitions that satisfy the
requirement of our definition and the final partition is related to
the initial conditions. For the karate club network, if we think the
club division is caused by some leaders, such as leaders (nodes)
$1,33,34$, and set the leaders and their random half neighbors as
initial partition, then our algorithm will divide the network into
$2$ communities. That is consistent with the real division. If we
set $1,2,33,34$ be the leaders, our algorithm will also partition
the network into $3$ communities which is the same as the result
without leaders. It is very interesting, with the process of group
formation, nodes $1,2$ and nodes $33,34$ combine and are in the same
community respectively. The other community don't contain any nodes
of $1,2,33,34$. It implies that, if some leaders have contradictions
and want to divide the network, some nodes will not always follow
the leaders and may form other groups (see Fig.\ref{ZK}).

\subsection{College football network}
We also apply our algorithm to Collage football network which was
provided by Newman. The network is a representation of the schedule
of Division I games for the $2000$ season. Nodes in the network
represent teams and edges represent regular-season games between the
two teams they connect. What makes this network interesting is that
it incorporates a known community structure. The teams are divided
into $12$ conferences \cite{GN}. Games are more frequent between
members of the same conference than between members of different
conferences. It is found that our overlapping algorithm identifies
the conference structure with a high degree of success. We detect
$12$ communities in which five communities were detected exactly,
the average accuracy is $0.8115$ and no node is overlapping. The GN
algorithm associating with Q function \cite{Newman evaluating
algorithm} gives the best partition with $Q$=0.2998. It divide the
football teams into $10$ communities and the average accuracy is
$0.6713$. The results are shown in Tab. \ref{Table football}.

\begin{table}
\centering{Table \ref{Table football}: The accuracy of each detected
community comparing with the counterpart of real-world
community.}\label{Table football}

\begin{tabular} [t]{|c|c|c|}
\hline
Conference name&Accuracy&GN accuracy\\\hline Atlantic
Coast&1&1\\\hline Big East&0.8000&0.8889\\\hline Big10&1&1\\\hline
Big12&1&0.9231\\\hline Conference USA&0.9000&0.9000\\\hline IA
Independents&0&0\\\hline Mid American&0.8667&0.8667\\\hline Mountain
West&1&0\\\hline Pac10&1&0.5556\\\hline SEC&1&0.7500\\\hline
Sunbelt&0.4444&0.4444 \\\hline Western Athletic&0.7273&0.7273
\\\hline Average accuracy&0.8115&0.6713\\\hline
\end{tabular}
\end{table}

\section{Conclusion and discussion}
In this paper, we present a new comparative community definition and
the corresponding algorithm. A community should satisfy that each
node's degree inside the community should be bigger or equal to the
node's degree toward any other community. Then we introduce the
concept of attractive force and develop a self-organizing algorithm
based on the comparing of attractive forces. The algorithm can
detect the community structures without any extra parameter. In
order to choose the best partition from several possible results, we
also define an indicator $H$ to evaluate the partitions. We apply
the algorithm to artificial networks and some real-world networks
such as Zachary karate club network and College football network.
The algorithm work well in all networks. Furthermore our community
definition and identification algorithm can be generalize to
weighted and directed networks easily.

Moreover, our algorithm can be use to predict network division when
there are some contradictions between some leaders. In the
algorithm, we can initially set some leaders and their random half
neighbors to be the communities respectively. Then the
self-organizing process gives a nice description of leaders'
affections. We think this partition technique has great potential
for analyzing network structure.

In section 2, we give the most weak community definition: in a most
weak sense, the sum of all degree inside the community should not be
smaller than the sum of degree toward any other community. From the
view of statical physics, we think the most weak definition is also
reasonable. Here we propose an open problem of finding a algorithm
to detect the communities based on the most weak definition.

\section*{Acknowledgement}
The authors thank Professor M.E.J. Newman very much for providing
College Football network data. This work is partially supported by
985 Projet and NSFC under the grant No.$70771011$, No.$70431002$ and
No.$60534080$.

\end{document}